\documentclass[aps,prd,twocolumn,groupedaddress]{revtex4}

\newif\ifpdf
\ifx\pdfoutput\undefined
\pdffalse 
\else
\pdfoutput=1 
\pdftrue
\fi

\ifpdf
\usepackage[pdftex]{graphicx}
\else
\usepackage{graphicx}
\fi
\usepackage{hyperref}

\begin{document}

\ifpdf
\DeclareGraphicsExtensions{.pdf, .jpg}
\else
\DeclareGraphicsExtensions{.eps, .jpg}
\fi

\def\hslash{\hbar}
\def\imag{i}
\def\grad{\vec{\nabla}}
\def\div{\vec{\nabla}\cdot}
\def\curl{\vec{\nabla}\times}
\def\DDt{\frac{d}{dt}}
\def\ddt{\frac{\partial}{\partial t}}
\def\ddx{\frac{\partial}{\partial x}}
\def\ddy{\frac{\partial}{\partial y}}
\def\lap{\nabla^{2}}
\def\divv{\vec{\nabla}\cdot\vec{v}}
\def\gradS{\vec{\nabla}S}
\def\vvec{\vec{v}}
\def\wc{\omega_{c}}
\def\<{\langle}
\def\>{\rangle}
\def\Tr{{\rm Tr}}
\def\Csch{{\rm csch}}
\def\Coth{{\rm coth}}
\def\Tanh{{\rm tanh}}
\def\g2{g^{(2)}}



\title{Interchain versus intrachain hole transmission through desoxyribonucleic acid molecular wires}
\author{Eric R. Bittner}
\affiliation{Department of Chemistry and Center for Materials Chemistry\\
University of Houston\\
Houston, TX 77204}

\date{\today}

\begin{abstract}
We present a methodology for 
computing the current-voltage response of a molecular wire 
within the Landauer-Buttiker formalism based upon transforming the 
cummulative transmission probability into an eigenvalue problem. 
The method is extremely simple to apply since does not involve construction of 
the  molecular Greens function, and hence avoids the use of complex
integration contours to avoid poles.  
We use this method to study the effect of base-pair sequence 
on the conductivity of holes in DNA chains containing A-T bridges between 
guanine chains.  Our results indicate that sequence 
plays a substantial role in ballistic transport  via 
tunneling resonances tuned by sequence and interchain interactions. 
We also find that ballistic transport is dominated by intrachain transport
and that hole transmission is insensitive to interchain fluctuations. 
\end{abstract}
\keywords{}

\maketitle

\section{Introduction}
Advances in synthetic technique and nano-scale device fabrication have facilitated the 
creation of prototypical electronic logic components in which the length scale 
of the device is on the order of the de Broglie wavelength of an electron.  
The prospect of using such  ``molecular electronic'' components in technological 
applications is a strong driving force in the emerging science and engineering of nanotechnology.  The real promise of nanotechnology relies upon on our
ability to predictably manipulate matter on the molecular length scale.  In particular, 
for molecular electronic materials, the goal is to be able to fine tune the quantum 
mechanical energy states so that the properties of the 
individual quantum states is expressed whatever macroscopic device we are
trying to engineering.   
The challenge here is then to build the interconnects between the 
individual molecules and the ``outside world'' so that one can reliably address
and access the molecular-scale components. 


The sequence dependence of charge transport in DNA is a topic of considerable
interest from both experimental and theoretical viewpoints.  
From a design standpoint,  DNA has a number of desirable properties. It 
possess  a high degree of of site specific 
binding between single strands of DNA and its related self-assembly properties.  
Furthermore, one can synthesize essentially any sequence desired and hence potentially
tune its properties in a controllable way.  
Such characteristics have made DNA an ideal candidate for incorporation into 
molecular scale  electronic devices.   With such motivation, a number of studies of the 
charge-transport properties of DNA have been performed
\cite{Lakhno04,Lakhno03,Bixon00,Bixon99,Voityuk01,Grozema99,Grozema00,Jortner02}
 and there is an emerging
consensus regarding the mechanism for single-electron transport in DNA chains.~\cite{DekkerRatner}
However, the DNA's intrinsic conductivity remains highly controversial.  Depending upon the 
experiment, DNA is a.) an insulator at room temperture~\cite{Braun98,dePablo00,Storm01,Zhang02},
b.) a wide-band gap semiconductor at all temperatures~\cite{Porath00,Rakitin01},
c.) Ohmic at room temperature and an insulator at low tempertures
~\cite{Fink99,Cai00,Tran00,Rakitin01,Yoo01,Hartzell03a,Hartzell03b},
or d.) metallic down to low tempertures~\cite{Kasumov01} with induced superconductivity.  
A recent review by Endres, Cox, and Singh nicely summarizes these varied experiments.~\cite{Endres04}

%


A number of  studies have been performed at various levels of 
theory with the aim of computing the current-voltage (IV) response of a  molecular 
wire.\cite{Ernzerhof03,Li01,Derosa2001,Damle2002,Brandbyge2002,Krstic2002,Magoga97,Emberly01,Zhang03,Zhang03,Nitzan,NitzanRatner,mujica1,mujica2,Datta,Datta2}
   The studies differ widely in terms of the level of treatment of the electronic
structure, the coupling between the molecule and the semi-infinite leads, and the change of electro-static
potential due to bias across the system.  Both first principles and semi-empirical treatments
have been used, with the former being restricted to smaller scale systems.

We consider the case where the terminal G's and $G-C$ pairs are 
anchored to opposing semi-infinite continua corresponding to the metal contacts. 
 Similar models have been considered by Jortner and Bixon~\cite{Jortner02,Bixon99}
considering the canonical transfer rate from $G_{i}$ to $G_{f}$ separated by $(T-A)_{n}$ bridges. 
Here, we consider the microcanonical transmission probability at a given scattering energy, $T(E)$. 
From this we can compute the transmission current as a function of applied 
voltage bias via\cite{Landauer,Buttiker,Nitzan,NitzanRatner,mujica1,mujica2,Datta,Datta2}
\begin{eqnarray}
I(V) = \frac{2e}{\hbar}  \int T(E,V) (f_L(E-V/2) + f_R(E+V/2)) dE
\end{eqnarray}
where $f_{L,R}$ are the Fermi distribution functions for the left and right hand
electrodes and $T(E,V)$ is the transmission probability at a given energy $E$  and 
applied voltage bias $V$.  
 We examine how permuting base-pairs changes the transmission 
probabilities and hence the current-voltage response. 
Aside from differences in parameterization of the models, 
the results presented herein parallel those of Roche~{\em et al.} who considered 
the the scaling behavior of the intrinsic conductivity of DNA due to 
length and sequence correlations.~\cite{Roche2003a,Roche2003b}  
Here we compare intra- vs. interchain hole transport whereas the work by Roche {\em et al} considered
considered the DNA strand as a single quasi-one dimensional strand.

Here present a simple approach for embedding a generic
quantum chemical problem into a scattering calculation for the purposes of computing 
the current-voltage response of molecular wire.   Our approach uses a 
idea originally developed by Manthe and Miller for computing the canonical thermal rate
constant for bimolecular reactions in reactive scattering.\cite{MantheMiller}
 The original idea was to apply absorbing boundary conditions about the transition state and use these to 
artificially impose the correct boundary conditions on the Greens functions used 
to construct the transmission matrix directly, thus, 
avoiding the explicit construction of the scattering matrix. 
In the case at hand, we impose absorbing 
boundary conditions through the use of a complex self-energy term for atoms  in our
system in direct contact with semi-infinite leads. 

\begin{figure}[t]
\begin{center}
\includegraphics[width=3in]{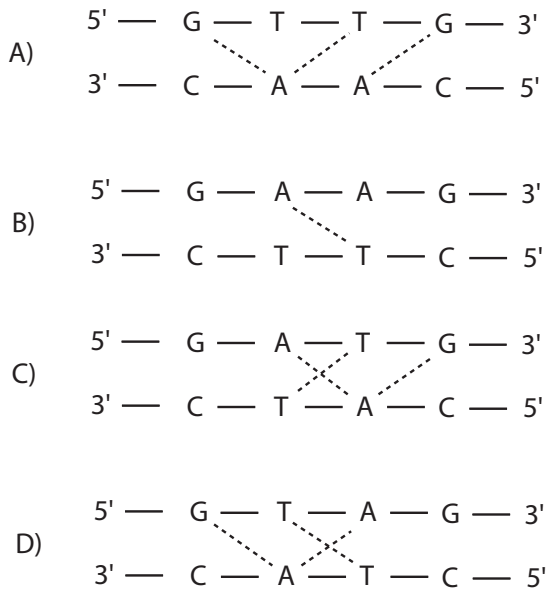}
\caption{Intrachain and interchain coupling pathways for 
hole transport in DNA sequences  considered. Solid connecting lines correspond to intrachain
hopping matrix-elements and dashed lined correspond to non-zero interchain hopping
matrix elements. }
\label{chains}
\end{center}
\end{figure}

%

\section{Hole transmission through model DNA sequences}
\subsection{Tight-binding model}
\begin{figure}[t]
\includegraphics[width=3.0in]{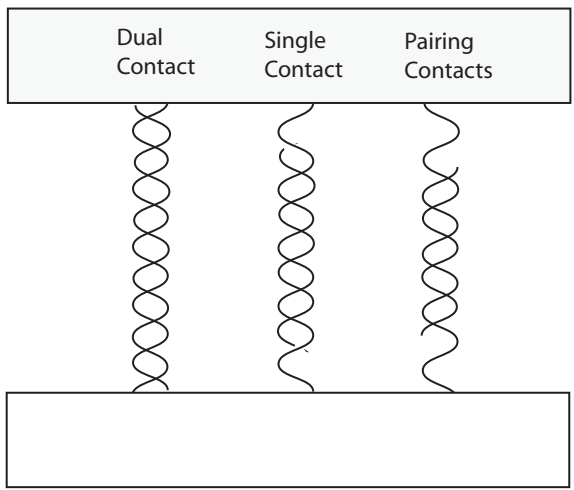}
\caption{Types of electrode-DNA-electrode contacts considered herein.  In the last case, 
complementary strands are connected to the opposing electrodes. }\label{configs}
\end{figure}

In what follows, we consider the bridging molecule to be a single DNA strand with matched
nucleotide pairs attached to the left and right metal clusters at the terminal 
sites as shown schematically in Fig.~\ref{configs}.   
Guanine, G, has the lowest oxidation potential of the four nucleotides and we consider the 
case where an injected hole travels along the DNA chain hopping on the sites 
containing G.  As the hole propagates along the chain, sites containing 
A-T pairs act as potential barrier between guanine sites that must be overcome by 
quantum mechanical tunneling.  Between guanine sites, hole transfer is taken to be 
effectively barrierless with no net change in free energy as the hole hops from $G_{n}$ to 
$G_{n\pm 1}$.  
Since scattering dynamics is dominated entirely by what happens 
when the particle encounters a potential change, the terminal $G$ units can be considered to be 
effective sites linking a short segment containing $(T-A)$ bridges to the asymptotic region. 
When there is a single barrier in for the hole, 
the hole can tunnel through the barrier and be transmitted.   If there
are multiple barriers, there is the possibility of resonant tunneling through 
quasi-bound hole-states on the chain.  These resonant states will appear as 
discrete peaks in the transmission probability as a function of energy.
Loosely speaking, each quasi-bound state corresponds to a particular single-particle
eigenstate delocalized over the entire chain. 

For this  consider a simple 
H{\"u}ckel chain connected to left and right electrodes at the left and right most sites:  
\begin{eqnarray}
H_m = \sum_{j} \epsilon_j c_j^\dagger c_j+  \sum_{i\ne j} t^{(m)}_{ij} (c_i^\dagger c_j + c_j^\dagger c_i) 
\end{eqnarray}
where $t^{(m)}_{ij}$ is the hopping integral, 
$c_j^\dagger$ and $c_j$
are fermion operators which create or remove a hole from the $j$th site,  
and $\epsilon^{(m)}_j$ the site energy.  For metal contacts, we constructed two $5\times 5 \times 5$ simple cubic 
lattices each located at the ends of the molecular chain.  
This we do to insure that the density of states of the contacts is properly
accounted for.  
The 	DNA sequence itself was connect to the 
center of the opposing faces of the left and right clusters.  
 The cluster Hamiltonians were taken to be 
tight-binding with 
\begin{eqnarray}
H_m = \sum_{j} \epsilon_{c,j} b_j^\dagger b_j+  \sum_{i\ne j} t^{(c)}_{ij} (b_i^\dagger b_j + b_j^\dagger b_i) 
\end{eqnarray}
where the hopping integral couples nearest neighbor sites.  Finally, the
clusters and DNA strands were connected via 
\begin{eqnarray}
V_{Lmc} &=& t^{mc}(b_{l}c^\dagger_j + b^\dagger_{l}c_j  )\\
V_{Rmc} &=& t^{mc}( b_{r}c^\dagger_k + b^\dagger_{r}c_k)
\end{eqnarray}
where the subscripts indicate the connected sites between the bridging DNA strand and the contact.

The various 4-base-pair chains considered 
here are depicted in Fig.~\ref{chains}. 
The dashed lines between chains
the indicate non-zero off-diagonal transfer integrals between sites on opposing chains. 
Each nucleotide site 
has associated with it a hole-wavefunction $|i\>$, a site energy, $\varepsilon_{i}$ 
and a transfer integral $t^{(m)}_{ij}$ to the other sites in the sequence.  
Since we are dealing with a tight-binding model, connectivity  
between sites plays a central role in determining the energetics of the system.  
We consider the following general scheme: For intrachain transport, hopping 
occurs between nearest neighboring pairs.  
For interchain hopping,
we assume that the hole on the $n$th unit can hop to an unit 
on the other chain one site removed, ${n\pm 1}$.   
Furthermore, we assume that interchain hops can occur between units of the same type, 
\begin{eqnarray}
\begin{array}{cl}
  term    &  t^{(m)}_{ij} \\
T_{n} \leftrightarrow T'_{n\pm 1}    &  (0.0032eV) \\
G_{n}\leftrightarrow G'_{n\pm 1}    &  (0.019 eV)\\
C_{n}\leftrightarrow C'_{n\pm 1}    &   (0.0007eV)\\
A_{n}\leftrightarrow A'_{n\pm 1}     &  (0.035 eV)
\end{array}
\end{eqnarray}
or between $G-A'$ and $A-T'$ interchain neighbors  
\begin{eqnarray}
\begin{array}{cl}
  term    &  t^{(m)}_{ij} \\
G_{n}\leftrightarrow A'_{n\pm 1} &(0.021 eV)    \\
A_{n}\leftrightarrow T'_{n\pm 1} & (0.016 eV)
 \end{array}
\end{eqnarray}
The logic behind this scheme is that there is better $\pi$ overlap between 
 interchain neighbors displaced by one step than those directly adjacent.  This allows 
relatively  facile hopping to occur between neighboring interchain pairs. 
 This  scheme has been verified via quantum chemical calculations by Voityuk\cite{Voityuk01} 
and the parameters for our model we take from a recent study by Lakhno, Sultanov, 
and Pettitt\cite{Lakhno04} who considered a combined hopping/super-exchange model for
computing hole-transfer rates through such model chains.  The values of the interchain hopping 
terms and remaining parameters are available as supplementary material. 
Since all of hopping terms are positive,  these can be considered as barriers for the 
delocalization of the hole over connected units.

\subsection{Embedding system in continuum} 
 
To connect the cluster-DNA-cluster system to a continuum, 
we consider the system as if it were embedded in between two 
semi-infinite continua acting as reservoirs.  
Once we assume this, we can partition the state space of
the total system into three domains, $Q_{L\alpha}$ and $Q_{R\alpha}$ which span the
 states of the left and right electrodes 
and $P$ which spans the electronic states of the bridging DNA molecule.
Thus, the full Hamiltonian has the structure
\begin{eqnarray}
H = \left(\begin{array}{ccc}
H_{L}  & V_{mL} & 0 \\
V_{mL}        & H_{m} & V_{mR} \\
0       &  V_{mR}     & H_{R} 
\end{array}
\right)
\end{eqnarray}
where the diagonal terms are the Hamiltonians for the uncoupled subsystems
and $V_{mL}$ and $V_{mR}$ are the couplings between the DNA and the 
left and right electrodes.    There is considerable ambiguity in how
this partitioning can be constructed since the metal contacts and the DNA bridge
are in intimate contact.  Here, we take a somewhat middle ground and define our partition between 
system and reservoir  by including the inner three layers of atoms in the contacts 
in the ``molecular'' subspace in order to properly include the surface density of states
and the electronic interaction between the DNA and surface in our calculations.  
Once we have made this partitioning, we can use the Schwinger-Keldysh formalism\cite{Joachain} to 
derive the self-energy operators, $\Sigma_{L}$ and $\Sigma_{R}$,  which embed the molecular 
sub-space into the continuum.  As a result, the Greens function for quantum particle moving
through the molecular subspace is
\begin{eqnarray}
G(E) = (E-H + \Sigma_{L}+\Sigma_{R})^{-1}.
\end{eqnarray}
Note that  in general $\Sigma_{L}$ and $\Sigma_{R}$ are complex non-local operators which depend upon the
scattering energy, $E$,
\begin{eqnarray}
\Sigma_{mm'} = \sum_{k}\frac{V_{mk}V_{km'}}{E-\epsilon_{k}+i\eta}.
\end{eqnarray}
Here, $V_{mk}$ is the coupling between the chain and the contact and the sum is over 
the energy eigenvalues $\epsilon_{k}$ of the contacts.\cite{Joachain}  The $i\eta$ in the 
denominator insures that the proper retarded scattering boundary conditions are enforced.  

Written in tight-binding site-representation, they can be written in a local form 
$$\Sigma_{L,n} = i\Delta_{n}(E) +  \Lambda_{n}(E)$$
where $\Lambda(E)$ shifts the site energy of the $n$th site.   $\Delta_{n}(E)$ has the effect of 
acting as an absorbing potential for the scattering wave function.  This insures   
that the scattering wave function obeys the proper asymptotic boundary conditions as we move 
away from the scattering region. 

We assume that the absorbing boundary conditions can be 
constructed as local self-energy terms within the subspace of the clusters
by setting the site energy of any atom located beyond a given cut-off radius from the 
point of contact between the bridge and the cluster to be complex
$$\epsilon_{c,n} = \epsilon  + i \Gamma$$
with $\Gamma = 0.2eV$. 
In the examples that follow, we set the cut-off radius such that at least two atomic layers 
of the contact cluster were inside the cut-off radius and two layers were outside the cut-off.  
This provided a reasonable compromise between having enough atoms in the cluster to 
represent the density of states of a simple cubic continuum, and provide a reasonable 
absorbing boundary condition region for the scattering wave functions.

\subsection{Computing Transmission and Current}

To compute transmission and current though the DNA strand, we use the Landauer-B\"uttiker formula
for relating the current, $I(V)$ to the transmission probability, $T(E,V)$, which 
may depend upon the applied bias, $V$, 
\begin{eqnarray}
I(V) = \frac{2e}{\hbar}  \int T(E,V) (f_L(E-V/2) + f_R(E+V/2)) dE \label{LBEquation}
\end{eqnarray}
where $f_{L,R}$ are the Fermi distribution functions for the left and right hand
electrodes.  The transmission probability is  microcanonical rate for an electron
injected on one end of the chain to leave on the other end as given by the Caroli equation,
~\cite{Caroli71,Joachain,SeidemanMiller,MantheMiller}
\begin{eqnarray}
T(E) = 4 Tr\left( \Gamma_{L} G_{r}^\dagger(E)\Gamma_RG_{r}(E)\right),
\end{eqnarray}
where $\Gamma_{L,R}$ are the imaginary parts of the left and right self-energies
describing the coupling of the molecular wire to the left and right hand continua.  
Finally, $G_{r}$ is the reduced  Greens function describing hole propagation through the 
bridging region 
\begin{eqnarray}
G_{r}(E) = (E-H + \Sigma )^{-1},
\end{eqnarray}
where $H$ is the molecular Hamiltonian with  $\Sigma = \Sigma_{L} + \Sigma_{R}$
being the self-energy operator of the hole on in the bridge.

We can also consider $I(V)$ in terms of the number of modes contributing to the current\cite{Datta}
\begin{eqnarray}
I = \frac{2e}{\hbar} M (F^+-F^-)
\end{eqnarray}
where $F^\pm$ are the potentials for the incoming and outgoing holes and $M$ is the number 
of modes.   We can generalize this by letting the trace in $T(E)$  be the sum over 
the probability that a given mode will contribute to the current at a given energy,
\begin{eqnarray}
T(E) = \sum_n p_n(E) = Tr(P).
\end{eqnarray}
Since the trace of a matrix is invariant to representation, we can construct the probability matrix,\cite{MantheMiller}
\begin{eqnarray}
P(E) = 4 \Gamma_L^{1/2} G^\dagger(E)\Gamma_RG(E)\Gamma_L^{1/2},
\end{eqnarray}
and $p_n(E)$ are then the eigenvalues of $P(E)$ which are bound by
$p_n(E)\in [0,1]$.   If we invert $P$, we have
\begin{eqnarray}
P^{-1} = \frac{1}{4} \Gamma_L^{-1/2}(E-H + \Sigma^*)\Gamma_R^{-1}(E-H + \Sigma)\Gamma_L^{-1/2}
\end{eqnarray}
and the eigenvalues of $P^{-1}$ are $p_n^{-1}$.  Consequently, we can compute the transmission 
and hence current by summing the eigenvalues of the $P$-matrix which we can construct
directly from $H$.  Any singularity which may be present in the Greens function no longer 
presents a problem when performing the energy integration in the Landauer-B\"uttiker equation. 

In practice, there is usually only one low-lying eigenvalue of the $P^{-1}$ matrix and 
this eigenvalue is typically on the order of unity. The remaining eigenvalues are typically
on the order of $10^{7}$ greater than the lowest eigenvalue.  Consequently, we can 
use matrix diagonalization techniques optimized for finding isolated eigenvalues.  
Since only the lowest eigenvalue of  $P^{-1}$ contributes to  
to  $Tr[P]$ to any significant extent, 
once this eigenvalue has been determined, we do not need to compute the remaining
eigenvalues.   Furthermore, for determining $N(E)$, we do not need to compute
the eigenvectors of $P^{-1}$.    Since computing $P^{-1}$ and its lowest eigenvalue 
involves simple matrix-vector manipulations, these tasks can be efficiently 
performed on a parallel computer.  Also, the method is universal and can be
used to compute transmission through any model system given model Hamiltonian and 
a self-energy representing the coupling of the model system to a continuum.  

One final comment regarding implementation of the Manthe-Miller algorithm is that
as expressed $P$ is a singular matrix if $\Gamma_{L}$ and $\Gamma_{R}$ are localized
to finite region and can not be formally inverted.  To remedy this, we allow the absorbing
potential to be {\em very small} in regions outside the absorbing region  $\Gamma_{L,R} = \eta $.
Typically, we set  $\eta$ to be between $10^{-7}$ to $10^{-10}eV$.

\subsection{Transmission through peptide chains}

We considered transmission through four model sequences as shown in Fig.~\ref{chains}.  
Furthermore, to explore the effect of intra- vs. interchain transmission we considered 
three possible contact scenarios between the chain and the electrode surfaces.  
In the first case, the terminal sites on both chains are connected directly to the cluster surfaces.  
In this case, hole injection and extraction occurs through both chains and can lead to 
destructive and constructive interferences as in a two-slit experiment.  Secondly, 
we considered the case where only one chain, the one with the terminal G sites, is 
directly connected to the clusters.  Here, the second chain acts as a hole acceptor and 
one expects to see only resonances coming from the intra-chain scattering pathways. 
Lastly, we consider the case where one chain is connected to  each cluster and the chains are 
connected by their complementary base pairs. This scenario has a potential technological ramification 
in designing current probes which bind only to specific complementary sequences on the opposite 
cluster surface.  In this case, current through the DNA must occur via interchain hops. This we 
expect to be quite weak, given that the interchain hopping matrix elements are roughly an order of
magnitude less than the intrachain hopping matrix elements. 

\begin{figure}
\includegraphics[width=3.5in]{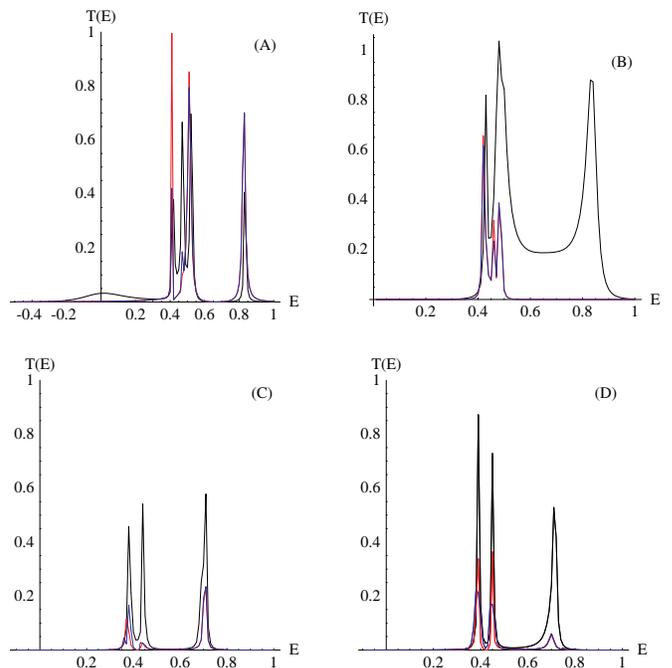}
\caption{(COLOR) Microcanonical transmission probabilities for hole transmission 
through various DNA sequences.  Black curves correspond to transmission 
with both terminal G and C groups connected to the surface of the electrodes, 
Red and blue curves correspond to only having the terminal G group on one strand 
attached to both electrodes.  For the blue curves, we randomly the interchain 
matrix elements to determine sensitivity to interchain interactions. 
 The units for the energy axis are eV.}\label{peptide-trans}
\end{figure}

The corresponding
transmissions through each sequence is shown in Fig.~\ref{peptide-trans}.  
In the black curves plotted in Fig.~\ref{peptide-trans}-A through D, both terminal G-C units are ``connected''
to the left and right metal clusters through transfer integral terms.  In this case, 
current can result via transfer through both opposing chains. 
The interchain hopping leads to constructive and destructive interferences
between the two current pathways. 
In the red curves plotted in  Fig.~\ref{peptide-trans}-A through D we consider the case in which only the 
terminal G sites are capable of transferring holes between the contacts and the chain.  
Here, loosely speaking, scattering interferences can occur when the hole hops from the initial chain
to the second and back to the first.    Since this process is also present in the 
dual contact cases, we can distinguish purely intrachain processes from multi-chain scattering 
processes by identifying peaks present in both spectra. 

In cases A and B we compare a $-A-A-$ bridge to a $-T-T-$ bridge.  In both cases, there is a 
weak broad resonance at $E = 0$.   However, the primary effect is to switch the order of the 
single resonance at $E  = 0.8 eV$ with the cluster of three resonances at $E = 0.5 eV$. 
It is also interesting to note that the gap between the single resonance and the 
cluster is almost unchanged by the switch.    Comparing the dual contact transmission to the 
single contact transmission, we note that  the single resonance peak is present in both 
in the single and duel contact cases 
and that two of the three resonances in the cluster about $E=0.5 eV$  are present in the single contact
case.  For B  the  resonance at $E \approx 0.4eV$ is present in both cases; however, 
in the single contact case, only one of the three corresponding resonances peaks from the duel contact case is present.   
In both A and  B,
  the weak resonance at $E = 0eV$ disappears when there is only a single 
contact. 

 In case C we have a $-T-A-$ linkage and we can see clearly three resonance peaks, a doublet centered 
 at  0.4 eV and a single peak at 0.7 eV.  In each of these cases, the corresponding eigenstate of the isolated
 molecular Hamiltonian is a delocalized interchain state.  The other eigenstates are more or less localized to 
 one chain or the other and do not extend fully across the bridge.    
Changing the DNA-metal contact so that only the terminal G's are connected profoundly 
diminishes the intensity of each transmission resonance.   Consequently, for the case of $-T-A-$ linkages, 
interchain pathways appear to play an important role
In case D, we transpose the linkage in case C to $-A-T-$.  Here, we observe the same 
resonance peaks as in case C, except with uniformly greater intensity even when we change the 
nature of the DNA-metal contact.  

%

Next, we consider the sensitivity of the transmission to interchain fluctuations.  
Within our model, the interchain hopping terms represent the tunneling splitting between 
two neighboring interchain sites.   Since tunneling is highly sensitive to distance of separation, 
structural fluctuations  should have a profound effect on the the ability of 
a hole to tunnel between the chains.   Such fluctuations can be included in the present model by 
sampling the interchain hopping matrix elements from a normal distribution centered about
some average value.  
For the blue curves,  we allowed 
each off-diagonal interchain term to fluctuate about its average value
 by $\sigma_{ j}= 0.01eV$  and then sampled over
an ensemble of 100 configurations.    Comparing the blue and red curves 
in  Fig.\ref{peptide-trans}A, the narrow resonance at 0.4eV is more or
 less washed out entirely by the interchain fluctuations. 
On the other hand, the two other resonances are clearly present in both the static case and in the 
fluctuating case indicating that at least some of the resonance features are insensitive to 
interchain fluctuations.  These surviving resonances are certainly due to intrachain transmission.   For the other three cases, the main resonance features are more or less
unaffected by the interchain fluctuations. 

In the last case, we connect one strand to the left 
contact and the complimentary strand on the other such that current can only pass through the
sequence through interchain  hopping.  Such a scenario could be useful in developing 
a sequence specific probe since current can only flow if the complementary strands match.
Unfortunately, however, our calculations indicate essentially zero transmission through 
any of the strand combinations considered above.    This is somewhat surprising considering that 
in each case at least one 
eigenstate of the isolated chain Hamiltonians 
are delocalized over the entire chain and have
some appreciable amplitude on both of the terminal contact sites.

\section{Discussion}
These results are both encouraging and discouraging for the potential utility of using current/voltage
probes to assay sequence.  They are encouraging since clearly sequence has a very profound impact on 
the hole transmission probability.  If one assumes that a given  chain can be decomposed into a series of
A-T potential barriers separated by a series of G's and that net transmission probability through 
the chain can computes as the product as a series of incoherent transition probabilities at a given energy,
one could effectively engineer a
DNA wire with specific voltage turn-on characteristics.   
For instance, having paired -A-A- or -T-T- sections 
as in 
$$5'-G-A-A-G-\cdots-G-T-T-G-3'$$
 would lead to a turn-on voltage 
of about 0.4 eV since both chain segments A and B have a series of resonances starting at this energy. 
Placing a -G-T-A-G- segment on the chain as in 
$$5'-G-A-A-G-G-T-A-G-G-T-T-G-3'$$
would lead to a slightly higher turn-on voltage since the transmission probability for the 
-G-T-A-G- (sequence D)
has a series of resonance peaks at $\approx$0.45 eV and this is in the middle of the 
cluster of resonances for sequences A and B.  
Moreover, transposing 
the A and T  to -G-A-T-G- (sequence C) as in
$$5'-G-A-A-G-G-A-T-G-G-T-T-G-3'$$
will result in a similar turn-on voltage for the current, however, the current will be roughly
half  that of when we have -G-T-A-G-.

What is also surprising is  that the transmission  is dominated by intrachain transport rather than 
a combination of intra- and interchain hopping.  
This is surprising given that the interchain hopping terms generally 
less positive then the intra-chain hopping matrix elements.  
A positive hopping term between sites implies that the {\em lower} energy
eigenstates will be the more localized than the higher energy states.  The greater the hopping term, 
the more localized the lower energy states become.  This is the opposite case 
from conjugated polyenes where the hopping term is negative leading delocalized low energy states
and resonance stabilization. 
Consequently, the delocalized interchain states should be lower in energy than the
delocalized intrachain states. The problem then is that they are insufficiently 
delocalized over the chain to provide a conduit from one end of the bridging molecule to 
the other.  

Finally, we note that the transmission only occurs above a certain threshold energy 
determined by the scattering resonances of the bridging system.  
Equally important  is density of available states in the regions on 
ether side of DNA bridge.   Here, we took the contacts to be three-dimensional simple-cubic 
solids with a fairly high density of states in the energies considered.  
However, in a real DNA molecular wire, the bridging region is more like
$$5'-G-\cdots-G-(\mbox{bridge})-G-\cdots-G-3'$$
and one should use a sequence of $G$'s in building the density of states and  the self-energies. 
If we consider a sequence of $5-G-\cdots-G-$ to be a quasi-one dimensional chain
with hopping integral $\gamma$ between neighboring sites, then the dispersion relation for the
chain is simply 
$$
\varepsilon_{k} = 2\gamma \cos(k)
$$
with $k$ in units of $h/2a$ and a being the lattice spacing.  Consequently, if the bandwidth 
$2\gamma$ is less than 
the resonance energy of the bridge,  the transmission through the bridge will not occur since 
the density of incoming states at that energy is zero.  On the other hand, if $\gamma$ is large, 
and one has a wide band for the holes in the guanine chains, then transmission through the bridge 
can occur.   This suggests that the conductivity properties of a DNA chain are highly dependent upon 
both the base-pair sequence of the DNA chain and the mobility of holes between neighboring $G$ sites.

\begin{acknowledgments}
This work was supported by the National Science Foundation and the Robert Welch Foundation.  
I also wish to acknowledge conversations with Prof. B. M. Pettitt concerning electronic transport in DNA 
sequences.
\end{acknowledgments}

\end{document}